\documentclass[a4paper,twocolumn,showkeys,floatfix,aps,pre,reprint,groupedaddress]{revtex4-1}
\usepackage{amsmath,amssymb}
\usepackage{graphics,graphicx}
\usepackage{dcolumn,bm}
\usepackage{psfrag}
\usepackage{enumerate}

\usepackage{xcolor}
\newcommand \David[1] {\bgroup\noindent[\textcolor{green}{\textbf{David}: #1}]\egroup\ignorespacesafterend}
\newcommand \MZ[1] {\bgroup\noindent[\textcolor{red}{\textbf{Michael}: #1}]\egroup\ignorespacesafterend}

\begin{document}

\title{Prediction of creep failure time using machine learning}

\author{Soumyajyoti Biswas${}^{1,2}$}
\email{soumyajyoti.biswas@fau.de}
\author {David F. Castellanos${}^{3}$}
\email{david.fernandez-castellanos@espci.fr}
\author{Michael Zaiser${}^{1}$}
\email{michael.zaiser@fau.de}
\affiliation{
${}^1$ WW8-Materials Simulation, Department of Materials Science, Friedrich-Alexander-Universit\"{a}t Erlangen-N\"{u}rnberg, Dr.-Mack-Str. 77, 90762 F\"{u}rth, Germany \\
${}^2$ Department of Physics, SRM University - AP, Andhra Pradesh, Guntur 522502, India \\
${}^3$ PMMH, ESPCI Paris/CNRS-UMR 7636/University Paris 6 UPMC/University Paris 7 Diderot, PSL Research University,
10 rue Vauquelin, 75231 Paris cedex 05, France
}

\date{\today}

\begin{abstract}
A subcritical load on a disordered material can induce creep damage. The creep rate in this case exhibits three temporal regimes viz. an initial decelerating regime followed by a steady-state regime and a stage of accelerating creep that ultimately leads to catastrophic breakdown. Due to the statistical regularities in the creep rate, the time evolution of creep rate has often been used to predict residual lifetime until catastrophic breakdown. However, in disordered samples, these efforts met with limited success. Nevertheless, it is clear that as the failure is approached, the damage become increasingly spatially correlated, and the spatio-temporal patterns of acoustic emission, which serve as a proxy for damage accumulation activity, are likely to mirror such correlations. However, due to the high dimensionality of the data and the complex nature of the correlations it is not straightforward to identify the said correlations and thereby the precursory signals of failure. Here we use supervised machine learning to estimate the remaining time to failure of samples of disordered materials. The machine learning algorithm uses as input the temporal signal provided by a mesoscale elastoplastic model for the evolution of creep damage in disordered solids. Machine learning algorithms are well-suited for assessing the proximity to failure from the time series of the acoustic emissions of sheared samples. We show that materials are relatively more predictable for higher disorder while are relatively less predictable for larger system sizes. We find that machine learning predictions, in the vast majority of cases, perform substantially better than other prediction approaches proposed in the literature.    
\end{abstract}


\maketitle


\section{Introduction}

All materials break under sufficiently high stress. However, even when the system can support a load at the instance of its application, it may still break at a later time by creep rupture \cite{andrade}. Local damage may accumulate even at a sub-critical loads. Accumulation of microstructural damage may be associated with the thermally activated crossing of energy barriers: examples include the accumulation of free volume as result of the thermal activation of shear transformations in disordered solids \cite{david18,david19}, or the thermally assisted removal of dislocation barriers in irradiated metals leading to microstructural slip localization and irradiation embrittlement \cite{zaiser2019_AEM}. Local damage accumulation reduces the energy barriers for future damage activation, thus promoting a tendency to localization. Overall, creep deformation is generally known to have three temporal regimes. First, we observe a decelerating strain rate regime associated with (statistical) hardening or aging effects as the weakest elements of the microstructure deform first and become consequentially inactivated by internal back stresses \cite{david19}. The decelerating regime is followed by an intermediate regime of constant strain rate and a final accelerating strain rate regime, associated with damage accumulation and strain localization and leading to catastrophic breakdown \cite{david18}. 

For obvious reasons, understanding the creep failure dynamics is an important issue for stability analysis of structures across scales. Especially, predicting the residual lifetime of a given sample until its failure under a subcritical load is a question that is  actively investigated by both physicists and engineers \cite{book}. Reliable lifetime predictions might not only avoid catastrophic in-service failure of components and systems, but also yield substantial economic benefits in view of the possibility of extending replacement cycles. Sample specific information on the damage accumulation process can, on the one hand, be obtained from the macroscopic sample response, i.e., the time dependent creep strain or strain rate. More detailed information can be drawn from analysis of the spatio-temporal pattern of energy releases as local creep damage accumulates in a material subject to subcritical load. The idea is here that the introduction of local damage is accompanied by a release of elastic energy which can be recorded by monitoring the acoustic emission (AE) of the sample, thus providing a means of non-destructively monitoring the damage accumulation process.

Among the empirical attempts to predict sample specific failure times from macroscopic creep strain rates, one possible approach is to correlate the time $t_{\rm m}$ of minimum strain rate with the catastrophic failure time $t_{\rm f}$, in the simplest case by assuming a linear relationship between both \cite{hao14,koivisto16}. However, there are multiple issues in using that observation for failure time prediction: (i) in analyzing time series for an individual sample, it is often difficult to identify a unique minimum for the strain rate. This problem is particularly pronounced when the creep strain rate is itself a stochastic, highly intermittent process (ii) While empirical observation indicates, on average, a linear relation between $t_{\rm m}$ and $t_{\rm f}$, the scatter is high especially for highly disordered samples. (iii) The prediction for $t_{\rm f}$ necessarily requires waiting until $t_{\rm m}$ can be reliably identified. Given that experimentally observed $t_{\rm m}$ already amount to $60\%$ of $t_{\rm f}$ and that larger times are needed to reliably identify a minimum, the resulting prediction might be too late to be useful \cite{alava2}. 

A different prediction approach focuses on temporal statistics of the damage accumulation process as monitored by AE. In this case, one looks at the magnitudes, times, and possibly locations of acoustic emission events and tries to identify statistical correlations that allow to interpolate the time of failure. For instance, one may exploit the observation made both in simulations \cite{david18} and experiments \cite{lennartz2014_PRE} that the AE event rate $\nu_{\rm AE}$ may accelerate towards failure according to a reverse Omori law, $\nu_{\rm AE} \propto (t - t_{\rm f})^{-p}$ with $p \approx 1$. Such a reverse Omori behavior was also reported to be a generic feature of mean-field models of thermally activated rupture processes \cite{saichev2005pre}. In such situations, one can obtain the failure time by fitting the Omori law to the AE record until time $t$, with the advantage that (unlike predictions based on the strain rate minimum) the ensuing predictions continually improve with increasing record length, i.e. decreasing time-to-failure. At the same time, the approach to failure may be accompanied with other characteristic changes in the AE burst statistics, such as an increase in the AE event size or characteristic changes in the Gutenberg-Richter exponent of the power law type energy statistics \cite{kun2013_PRE,david18}, which may also be used for monitoring and prediction purposes. 

Even further information can be harnessed by simultaneously monitoring the spatial pattern of damage accumulation activity, as failure is associated with localization of damage \cite{kun2013_PRE,lennartz2014_PRE,david18}. Spatio-temporal correlations in energy release signals, therefore, could hold important information regarding distance to the catastrophic breakdown of the sample. However, given the high dimensionality of the data sets involved and the possible complexity in the correlation measures, it may not be possible to extract the necessary information regarding failure time in terms of simple empirical laws. The task of extracting non-trivial correlations from high dimensional data is precisely what machine learning algorithms can do best. Indeed, in recent times, machine learning found widespread applications in predicting deformation, failure, and flow processes in disordered systems \cite{labquake,salm18,geophys}. Here we use Random Forest regression\cite{ml_book} for extracting information regarding sample specific failure times from spatio-temporal records of energy release signals prior to failure. To avoid problems resulting from scarcity of data, we obtain our training and testing data from ensembles of creep rupture simulations performed using the model introduced in Ref. \cite{david18}. The trained algorithm is tested over a set of samples previously unseen by the algorithm using various accuracy measures. We investigate the variations in prediction accuracy as a function of loading shear stress, the degree of microstructural disorder, and sample size.

\section{Results}

As a model for creep rupture, we use a mesoscale elastoplastic model \cite{david18,david19} that considers plastic activity accompanied by damage accumulation in a simulated sample which is driven by a temporally constant, subcritical shear load (see Methods). The sample volume is divided into mesoscopic volume elements. Local energy barriers control deformation and damage accumulation within the individual elements. The statistical distribution of these barriers characterizes the microstructural disorder of the material. The barrier height is reduced by stress, hence, if local stresses are high enough, barriers may be crossed and local plastic activity takes place. At the same time, internal stresses which arise from local deformation couple the deformation response of the individual elements. Plastic deformation generates local damage which reduces, on average, the local barrier height. The coupling between deformation, internal stresses and damage accumulation ultimately leads to damage localization in the form of a macroscopic shear band. Such damage localization induces a divergence of the strain rate which indicates catastrophic failure. The model has been successful in reproducing the temporal regimes of creep, the statistics of activity in the form of avalanches and progressive strain localization \cite{david18,david19}. A detailed model description and default model parameters are provided in the Methods section.

\begin{figure}
\centering
\includegraphics[width=8cm, keepaspectratio]{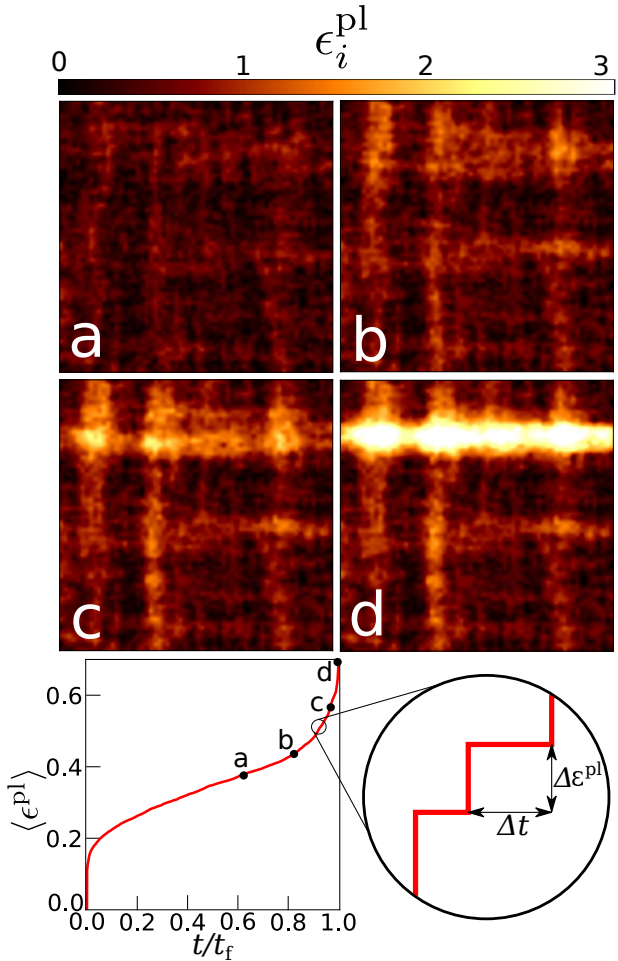} 
\caption{The top panel shows the spatial evolution of damage (cumulative number of local AE events) as time evolves 
($t(a)<t(b)<t(c)<t(d)$). At later time, the damage becomes localized. The bottom panel shows the growth of the global AE event number with time. The qualitative signature of localization appears roughly above $0.7t_{\rm f}$, which is already close to breakdown. The aim of this study is to make predictions substantially ahead of the manifestation of damage localization.}
\label{drawing}
\end{figure} 

The model produces, as raw data output, information that can be interpreted as a simulated Acoustic Emission time series: Deformation activity is characterized by the timings, locations, and amplitudes of deformation avalanches, as well as by the resulting spatio-temporal strain patterns. As shown in \cite{david18,david19}, the corresponding time series exhibit correlations which evolve with time. Examples are the variation of the statistics of avalanches or the progressive localization of spatial activity (see Fig. \ref{drawing}). These variations depend on the proximity to failure and can thus be envisaged as precursors with the potential for prediction. As explained in the Methods section, from the time series of the mesocale creep simulation we extract several features that contain such information. At each time $t$ measured since the beginning of the creep process, the machine learning algorithm makes a prediction for the remaining time to failure, $t_{\rm p}$. To each time $t$ at which a prediction is made, we can {\em post mortem} assign an actual remaining time to failure $t_{\rm a}$. Therefore, we define the fractional error of the machine learning prediction as $e_{\rm ML}=\frac{|t_{\rm p}-t_{\rm a}|}{t_{\rm f}}$. 

The quality of the machine learning predictions is highly dependent on the size of the training set. Consequently, it is convenient to quantify the learning performance of the algorithm not in absolute terms but relative to what one would predict without a machine learning algorithm, for example by a simple mean of the training set. To this end, we compute an average remaining time to failure over the training set and assume that each test sample would simply follow such average. The error made by such prediction is denoted by $e_{\rm woML}$. We define the relative improvement achieved by machine learning over a naive average prediction as $\epsilon = e_{\rm ML}/e_{\rm woML}$.

We use the creep time series of 1000 samples as the training set and 200 different samples as the test set to evaluate the predictions of remaining time to failure. With the trained algorithm, we systematically study variations in the prediction accuracy measures mentioned above for different values of the external applied stress, disorder and sample size. Afterward, we benchmark the machine learning predictions against other methods of prediction that use empirical laws. Specifically, we use the time minimum of the strain rate $t_{\rm m}$ to predict the failure time $t_{\rm f}$ assuming a linear relationship between both.  

\subsection{Dependence of prediction performance on applied stress level}

In order to reproduce creep conditions, the system is loaded with a constant external stress $\Sigma^{\rm ext}$ which is below the short-term critical stress $\Sigma^{\rm c}$ at which the system fails instantaneously. The failure time $t_{\rm f}$ depends strongly on the ratio $\Sigma^{\rm ext}/\Sigma^{\rm c}$. It is therefore natural to expect a variation in the predictability as $\Sigma^{\rm ext}/\Sigma^{\rm c} \to 1$. We have used three values of applied stress -- 60\%, 70\% and 90\% of the critical stress, respectively, keeping the other parameters of the model fixed. 

\begin{figure}
\centering
\includegraphics[width=8cm, keepaspectratio]{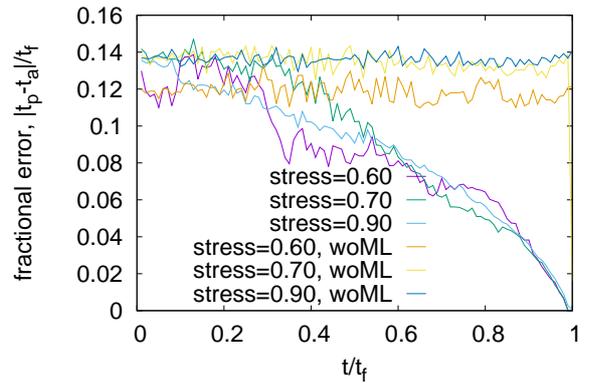} 
\caption{The fractional error in ML predictions at various stress levels and the corresponding predictions based upon average failure times. The two predictions start from similar levels and the ML predictions improve gradually, while the other prediction remains unchanged. There is no systematic dependency of prediction accuracy on stress level.}
\label{avg_error_stress}
\end{figure}

Fig. \ref{avg_error_stress} shows the fractional errors $e_{\rm ML}$ and $e_{\rm woML}$ for different values of stress, as mentioned above. Even for this considerable range of variation in the applied load, we find no systematic dependence of the prediction score on stress. It is interesting to note that for small values of $t$, the prediction from machine learning is just equal to the average of the training set. This is expected since, at the beginning of the creep dynamics for a particular sample, the algorithm has not yet received any sample specific information. As time progresses, the algorithm utilizes its training and makes, based on the precursor activity up to that time, predictions that improve with increasing length of the precursor record. On the other hand, in the absence of any `training', the naive prediction from the average of training data does not improve with time and remains roughly constant.

\begin{figure}
\centering
\includegraphics[width=8cm, keepaspectratio]{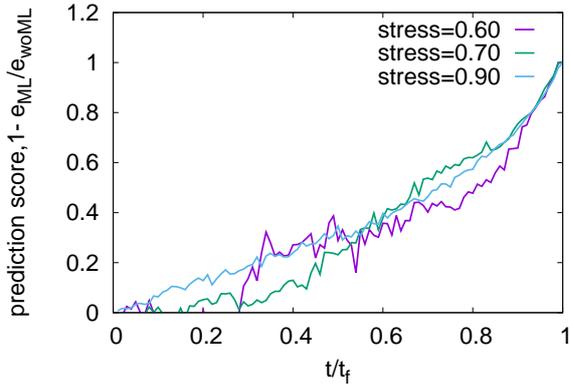} 
\caption{Prediction scores as function of time to failure, for different applied stress levels. See Methods for model parameters.}
\label{avg_relative_error_stress}
\end{figure}

In the following we define the prediction score in terms of the actual prediction error, divided by the prediction error of a naive average prediction. A prediction which is just as good as the average lifetime in the training ensemble is assigned a prediction score of zero, and an exact prediction of the failure time achieves a prediction score of 1. We thus set
\begin{equation}
S = 1 - \frac{e_{\rm ML}}{e_{\rm woML}}.
\end{equation}
Fig. \ref{avg_relative_error_stress} shows the prediction score achieved by machine learning for different stress levels, as a function of time to failure. Note that the extreme increase of the damage rate just before failure ensures that failure is always correctly identified as it happens, with the consequence that for $t \to t_{\rm f}$, $S \to 1$. The question is, however, whether the machine learning algorithm can achieve good prediction scores at earlier times. 

\subsection{Dependence on material disorder}
In the model we use, the local barrier heights which control damage accumulation are statistically distributed to represent a material with a disordered microstructure. If one assumes the weakest-link hypothesis, then the local strength of a mesoscale region is essentially the strength of the weakest microscopic subregion. In this case, the mesoscale distribution function of the local strength is expected to follow a Weibull distribution \cite{david19}. We statistically distribute the local barriers according to a Weibull distribution with shape parameter $k$, which determines the width of the distribution and hence can be used to quantify the microstructural disorder. Specifically, a small value of $k$ indicates a wide distribution and therefore a high degree of microstructural disorder. This translates into a comparatively large statistical scatter of the sample lifetimes. Conversely, very large values of $k$ imply nearly deterministic behavior, i.e., the creep curves of different samples and the corresponding sample lifetimes are almost identical. 

\begin{figure}
\centering
\includegraphics[width=8cm, keepaspectratio]{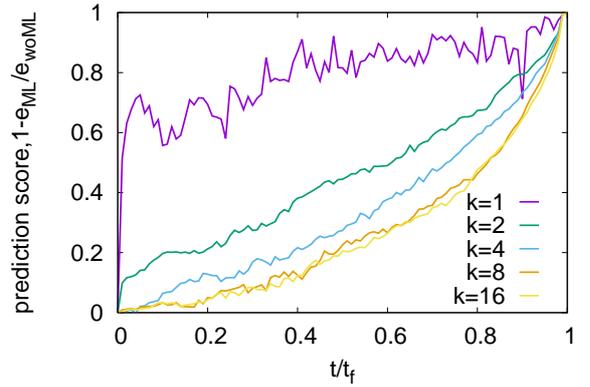} 
\caption{ML prediction scores for different degrees of disorder (Weibull exponents $k$), see Methods for model parameters. High prediction scores can be achieved even after short creep times for materials with a high degree of disorder. For very low disorder, 
the prediction scores decrease and ultimately saturate.}
\label{avg_relative_error_disorder}
\end{figure} 

Fig. \ref{avg_relative_error_disorder} shows time dependent prediction scores achieved for a range of values of $k$. High disorder substantially improves predictability of the remaining time to failure. The reason for the higher prediction scores lies in the more complex precursor activity and larger variations in local properties, resulting in statistical correlations which anticipate strain localization already at early creep stages well before catastrophic failure (see e.g., \cite{vasseur}), leading to better-trained algorithms for the same number of training samples.

On the other hand, with decreasing disorder prediction scores become independent of the disorder. This can be understood by noticing that when the disorder distribution is very narrow, the stochastic behavior of the time series becomes dominated by thermal noise, which is kept constant. Hence the predictability becomes independent of the disorder of local strengths.

\subsection{Dependence on sample size} 

The mesoscale model considers a square lattice composed of $L \times L$ mesoscale regions. Here we vary the linear system size $L$ to study the impact of sample size on the accuracy of the machine learning predictions.

\begin{figure}
\centering
\includegraphics[width=8cm, keepaspectratio]{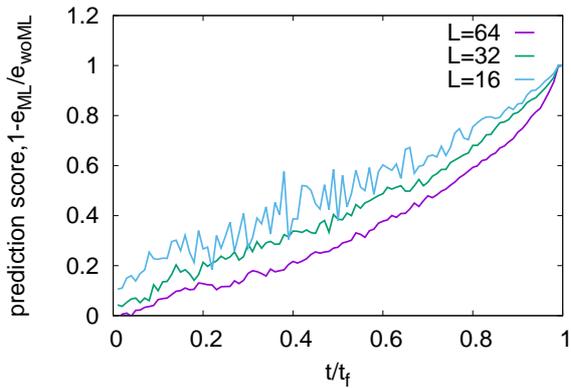} 
\caption{Dependency of ML prediction scores on system sizes, see Methods for model parameters: smaller systems are more predictable.}
\label{avg_relative_error_size}
\end{figure}

As can be seen from the plots in Fig. \ref{avg_relative_error_size}, the prediction scores decrease with system size. This can be understood as a consequence of strain localization. Sample failure is controlled by processes taking place in a localized shear band which emerges before failure. The width of this band does not depend on system size, hence it occupies a smaller fraction of the sample if the sample is larger. If one assumes that precursory signals that can be used for prediction mainly emanate from the shear band region (see \cite{lennartz2014_PRE} for a discussion of this phenomenon on a real sample), whereas other regions mainly produce confounding 'noise', then it is clear that smaller samples exhibit a better ratio of precursory signal to stochastic noise, and are therefore more predictable. 

This observation might have far-reaching consequences in terms of real-world predictions. For example, a catastrophic shear band in a laboratory-scale fracture test of a rock sample occupies a far larger fraction of the overall sample volume than the slip localization zone in the context of an earthquake. Thus, sample size may be an important factor determining predictability, with unfortunate implications for the predictability of geo-scale fracture processes.  

\subsection{Comparison of ML with alternative prediction methods}

\begin{figure}
\centering
\includegraphics[width=8cm, keepaspectratio]{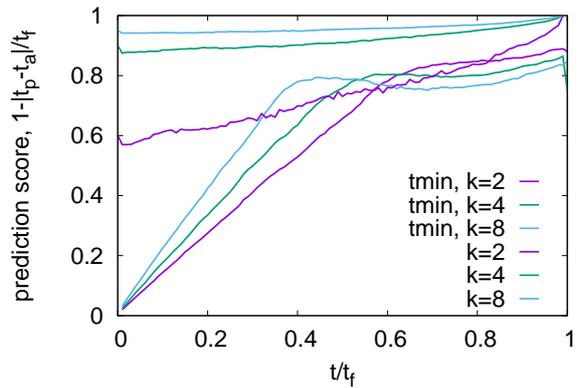} 
\caption{Comparison between the prediction scores from machine learning and failure time prediction based upon minimum creep rate. 
See Methods for model parameters. }
\label{avg_relative_error_tmin}
\end{figure}

Finally, it is useful to compare the machine learning predictions done here with other methods. We choose an empirical method which uses the time $t_{\rm m}$ at which the global minimum strain rate occurs to linearly extrapolate and predict failure time $t_{\rm f}$ \cite{hao14,koivisto16}. The main drawback of this method lies in the fact that the average minimum is quite flat, whereas the instantaneous creep strain rate is subject to strong fluctuations. To apply this method, a smoothed  signal must be constructed first from the discrete sequence of events (see Methods). The question is whether the best possible prediction from the empirical method based on the global minimum is better than the one from machine learning. Fig. \ref{avg_relative_error_tmin} shows such a comparison. As expected, the initial predictions from the strain rate minimum are, during the initial stages of creep before the actual minimum has been reached, far worse than the machine learning ones. The same is true during the late stages of creep close to the failure time, since predictions based on the strain rate minimum cease to improve once the minimum is passed, whereas ML predictions continuously improve. For moderate to low disorder ($k=4-8$), the strain rate minimum method performs consistently worse than the machine learning method. Only for high disorder ($k=2$) and creep times close to $t_{\rm m}$ it achieves prediction scores that are comparable to ML. 

\section{Discussion and conclusions}

Prediction of failure time for creep rupture is a crucial problem with  wide-ranging potential applications in science and engineering. Empirical prediction methods are often not accurate enough, especially when the disorder is strong. Moreover, sample to sample variations are often high which makes it difficult to extrapolate knowledge gathered from a tested sample to another one. For this reason, we have trained a machine learning algorithm. For machine learning, fluctuations become a source of knowledge that can help in training the algorithm to better recognize precursor patterns to failure and exploit complex correlations which can be used to predict incipient failure.

We have performed a systematic study of the variations of predictability with applied stress, presence of disorder and sample size, using synthetic data generated from a well-established model of creep deformation and failure. We find no systematic variation of predictability with stress. However, predictability increases with increasing disorder while it decreases with increasing sample size. Our benchmark against alternative methods confirms a superiority of machine learning over other approaches suggested in the literature, which can be regarded as a promising method with the potential to improve existing hazard assessment techniques. 

The proposed method is based on the use of a set of features characterizing the deformation process which finds a direct equivalent in laboratory tests, such as the variation of the intensity of acoustic emissions or signatures of progressive spatial localization. Advances in acoustic emission tests have made possible in recent years to measure both magnitudes with a high resolution, which makes the proposed machine learning method an ideal candidate for analyzing and extracting useful information from experimental data.

\section{Methods}

\subsection{Creep model}
Synthetic space-time series of creep deformation accompanied by damage accumulation are produced by a mesoscale model of plastic deformation of disordered materials introduced in \cite{david18,david19}. The model considers a 2D $L \times L$ lattice of mesoscale elements denoted by an index $i \in [1\dots L^2]$. Each mesoscale element has a volume $V$ which coarse-grains microscopic details of a disordered material. We describe the state of each mesoscale element by continuum mechanics variables, namely a tensorial irreversible strain $\boldsymbol{\epsilon}_i^{\rm pl}$ and a stress tensor $\boldsymbol{\Sigma}_i$ which is connected to the reversible (elastic) strain tensor via the tensor of elastic constants, which we assume to represent an isotropic material. The internal microstructure of each element is characterized by a spectrum of stress dependent energy barriers of which we assume the lowest barrier, $\Delta E_{{\rm min},i}(\boldsymbol{\Sigma}_i)$ to control activation of irreversible deformation. To make the connection with traditional concepts of mechanics of materials, we introduce element specific, stress dependent yield functions $\Phi_i(\boldsymbol{\Sigma}_i)$ which fulfil the condition 
\begin{equation}
\Delta E_{{\rm min},i}(\boldsymbol{\Sigma}_i) = 0 \quad {\rm if} \quad \Phi_i(\boldsymbol{\Sigma}_i) = 0
\end{equation}
We assume that deformation is controlled by deviatoric (shear) stress only and take $\Phi_i$ to be of the form
\begin{equation}
\Phi_i =  \hat{\Sigma}_i - \Sigma^{\rm eq}_i
\end{equation}
where $\Sigma^{\rm eq} = \sqrt{(3/2) \textrm{dev}(\boldsymbol{\Sigma}):\textrm{dev}(\boldsymbol{\Sigma})}$ is the von Mises equivalent stress and $\textrm{dev}(\boldsymbol{\Sigma})$ denotes the deviatoric stress tensor. $ \hat{\Sigma}_i$ defines the equivalent stress at which $\Phi_i=0$, i.e., the stress at which the energy barrier to initiate a plastic strain increment vanishes and, hence, the volume element $i$ becomes mechanically unstable. In the language of plasticity theory, this corresponds to the local flow stress in the limit of zero temperature. 

In the regime of negative $\Phi_i$, plastic deformation is controlled by thermal activation over non vanishing barriers. For simplicity, we assume the rate controlling energy barrier to be linearly proportional to $\Phi_i$, i.e., 
\begin{equation}
\Delta E_{{\rm min},i}(\Phi_i) = \left\{\begin{array}{ll}
- V_{\rm a} \Phi_i \quad,& \Phi_i < 0,\\
0 \quad, & \Phi_i \ge 0. 
\end{array}\right.
\end{equation}
Barrier crossing leads to a discrete plastic event which introduces a finite tensorial plastic strain increment $\Delta \boldsymbol{\epsilon}^{\rm pl}$. The barrier crossing rate in element $i$ is $\nu_i =  \nu_{\rm el} \exp(-\Delta E_{{\rm min},i}/k_{\rm B}T) = \nu_{\rm el} \exp(\Phi_i/\Sigma_T)$ where the parameter $\nu_{\rm el}$ defines the local yielding attempt frequency at the mesoscale and $\Sigma_T = k_{\rm B}T/V_{\rm a}$ characterizes the influence of temperature in terms of a scalar, stress-like variable. In mechanically unstable elements ($\Phi \ge 0$) events are assumed to occur instantaneously. Upon activation of an event in an element, the plastic strain field in that element is updated by adding to the local irreversible strain the tensorial increment $\Delta\boldsymbol{\epsilon}^{\rm pl}_i = \Delta\epsilon^{\rm eq}_i \cdot \boldsymbol{\hat{\epsilon}}_i$, where $\Delta\epsilon^{\rm eq}_i$ denotes the scalar magnitude of the strain increment and the tensor $\boldsymbol{\hat{\epsilon}}_i$ defines the direction. In line with J2 plasticity, this direction is given by a maximum energy dissipation criterion, thus $\boldsymbol{\hat{\epsilon}} = \nabla_{\boldsymbol{\Sigma}} \Phi = (3/2)\textrm{dev}(\boldsymbol{\Sigma})/\Sigma^{\rm eq}$. On the other hand, the magnitude is given by $\Delta\epsilon^{\rm eq} = \chi\Sigma^{\rm eq}/3G$, where $\chi$ is a factor between 0 and 1. This choice ensures that the local deviatoric stress (the thermodynamic driving force) cannot change sign upon introduction of an event. 
The location $i$ of a thermally activated deformation event and the associated time increment $\Delta t > 0$ elapsed since the last thermal activation are in our simulations determined by the Kinetic Monte Carlo method. Upon introduction of an event in element $i$, we increase the plastic strain tensor in that element, $\boldsymbol{\epsilon}^{\rm pl}_i \to \boldsymbol{\epsilon}^{\rm pl}_i + \Delta\boldsymbol{\epsilon}^{\rm pl}_i$. Alongside with the plastic strain tensor, we also update the cumulative equivalent strain, $\epsilon^{\rm eq}_i \to \epsilon^{\rm eq}_i + \Delta\epsilon^{\rm eq}_i$. Using the updated plastic strain field, stresses everywhere in the simulated sample are re-computed. The ensuing stress changes may lead to destabilization of other elements and thus to secondary events. In that case, stresses are again updated considering all such plastic events simultaneously, then checking for further unstable events, and continuing this cycle until the system is mechanically stable and the 'avalanche' terminates. The simulation then returns the following primary data: (i) the time of the thermally activated event, (ii) the overall strain increment, (iii) the change in the spatial strain pattern. 

To account for microstructural randomness, the element strength $\hat{\Sigma}_i$ is statistically distributed according to a Weibull distribution of exponent $k$ with cumulative distribution function
\begin{equation}
P(\hat{\Sigma}_i) = 1 - \exp\left(-\left[\frac{\hat{\Sigma}_i}{\bar{\Sigma}_i}\right]^k\right)
\label{weibull}
\end{equation} 
and mean value $\langle \hat{\Sigma}_i \rangle = \bar{\Sigma}_i \Gamma(1+1/k)$ where $\Gamma$ denotes the Gamma function. Whenever an element undergoes plastic deformation, its strength is renewed from the distribution (\ref{weibull}). Deformation-induced damage in element $i$ is described by a variable $\delta_i = f \epsilon^{\rm eq}_i$ which is proportional to the local equivalent plastic strain. The average of the distribution from which local strength values are drawn decreases with local damage as $\bar{\Sigma}_i = \bar{\Sigma}_0\textrm{exp}(-\delta_i)$, implementing strain softening.

We load the system under pure shear conditions, with principal axes oriented along $\pm \pi/4$ to the $x$ axis of our Cartesian coordinate system. This gives rise to a spatially homogeneous 'external' stress tensor which represents a pure shear stress state, $\boldsymbol{\Sigma} = \Sigma^{\rm ext}({\bf e}_x \otimes {\bf e}_y + {\bf e}_y \otimes {\bf e}_x)$. In the course of creep deformation, the emerging inhomogeneous plastic strin patterns leads to inhomogeous and multi-axial internal stresses which add to this external stress. The calculation of these stresses is done by the Finite Element Method with a regular square grid, linear shape functions, assuming linear elasticity and with homogeneous elastic properties. Each mesoscale element is matched with a finite element. The stress at each mesoscale element is computed as the average of the stress field within the associated FEM element. The reader is referred to \cite{david_thesis} for further details on the numerical implementation of the model.

To perform simulations under creep loading conditions, the value of $\Sigma^{\rm ext}$ is kept fixed in time. To establish the specific value, we look first for the critical value $\Sigma^{\rm ext}=\Sigma^{\rm c}$ beyond which the system is mechanically unstable and fails instantaneously even at zero temperature. This parameter defines our stress scale. Accordingly, we measure stresses in units of $\hat{\Sigma}(0)$, strains in units of $\hat{\Sigma}(0)/E$ where $E$ is the Young's modulus of the material, and time in units of $\nu_{\rm el}^{-1}$. Externally controllable parameters are $\Sigma^{\rm ext}$ and $\Sigma_T$. Default parameters in our simulations are, unless otherwise stated, 
$L=64$, $\Sigma_T=0.0075$, $\chi = 0.05$, $f=0.1$, $k=4$ and $\Sigma^{\rm ext}=0.7\Sigma^{\rm c}$. 

The model produces, as output, raw data in the form of the times, locations, and magnitudes (strain increments) of all local deformation events between initial loading and failure. We can envisage this output as a simulated acoustic emission record which monitors the deformation activity within the sample throughout the creep process. 

\subsection{Machine learning method}

We use a supervised learning algorithm - Random Forest - for predicting the failure time. The algorithm is trained over a training set where we use various features of the creep simulation data points. Predictions are made at the times where an avalanche is ended. While for making a prediction at a given point of time (necessarliy at the end of an avalanche) we use the data features fed to the algorithm at that point of time, it is strictly speaking not only the instanteneous information, since the features of the data series include both temporally and spatially aggregated information upto that point.   Specifically, the features used for prediction include (1) the elapsed time since the beginning of the creep process up untill the point where a prediction is being made (again, necessarily at the end of an avalanche), (2) the size of the last avalanche (at the end of which the prediction is being made). We also add spatially-aggregated information such as the (3) maximum and (4) minimum damage (accumulated local AE events as shown in Fig. \ref{drawing} (a)-(d)) magnitudes. Attributes (3) and (4) are calculated simply by taking the sum over the individual rows and columns of the matrices shown in Fig. \ref{drawing} (a)-(d), and noting the \textit{magnitude} of the maximum and minimum accumulated damage along a row \textit{or} column, i.e. if $d_{ij}$ represent the accumulated AE damage matrix, then let $D^{row}_{max}=max(\sum\limits_i d_{ij}, j=1,2, \dots, L)$ and $D^{col}_{max}=max(\sum\limits_j d_{ij}, i=1,2, \dots, L)$, then the third attribute is simply $max(D^{row}_{max},D^{col}_{max})$. The fourth feature is just the corresponding minimization problem. Given the direction of the applied shear, the shear bands are along either of the axes (and not diagonal or along any other angle). Hence the sums along row and columns of the accumulated damage matrix ($d_{ij}$) are able to assess the most vulnerable damage. Finally, the last column of the training data set is the target variable, i.e. time remaining before macroscopic failure, which we want to predict (in the test data set). 

For every prediction set, a typical training data set is the above mentioned features of 1000 samples and the test data set is typically of 200
samples (over which the prediction accuracy is measured). The training set data file is the combination of the all 1000 samples time series features
appended together. Due to the slow time variation and the size of file, one in 50 time steps are considered for the training set to have a 
significant change in the feature values.  Further details about the data processing and parameter sets for the 
model of regression are given below.

\subsubsection{General features of the Random Forest regression model}
The algorithm used here for machine learning is Random Forest (RF) regression \cite{ml_book}. It is an ensemble algorithm that
makes predictions based on the average prediction of an ensemble of decision trees. A decision tree is a flow-chart like
structure, where starting from a root node, the samples are split depending on their feaure values or attributes. 
For example, a particular attribute $A$, could be used to split the samples into two parts, those having values
less than $A_0$ and those having values greater than $A_0$. Each of these parts can be further split depending
on the other attribute values and so on. The spliting values of the attributes at each stage are optimized by the algorithm used until 
all samples at a given node has the same value of the target
variable (in this case, time to failure) or, further spliting does not improve predictions (fixed by the variance reduction criterion \cite{var_red} or fixed by an upper cut-off in the number of possible spliting i.e. the \textit{depth} of the tree). The end nodes are called leafs and
they hold the predictions for the given set. Now, for each of the trees, the training data are subsampled using a Bootstraping 
algorithm (see below). Consequently, each tree is fed with a random subset of the training data (hence Random Forest).  Following
the training, the test data, which is unseen by the model until this point, are passed through each tree and they end up in the leaf nodes
which are then the predictions for each of the test data points. In case of regression, where the target variable is continuous, like here, 
the prediction of the RF for a given test data point, is the average value of the prediction of each of the tree for the same test data point.

\subsubsection{Data processing for regression}
Here, from the training data set, a number of subsampled data sets
are generated by randomly selecting data points from the training set (selecting, say, $N$ rows randomly and uniformly from $N$ rows, 
but with replacement i.e. bootstrapping). 
 Due to bootstrapping, some of the data points will
be repeated, which acts as mitigation towards outliers in the training set. The number of subsampled, randomly generated, 
sets is equal to the number of decision trees used in the RF (see below). Each of the trees are then fed with a different training set
(randomly sampled) and in the case of regression, as in our case, the average prediction of all the trees is the prediction of the forest.

\subsubsection{Parameters of the model}
As mentioned above, the RF algorithm uses a set of decision trees for making predictions. The number of decision trees used here is 1000. 
The algorithm is also specified with the maximum depth (number of spliting) for each tree, set at 10. Higher depth leads to increase in error due to overfitting. 
The minimum samples required to split an internal node is 2 and the minimum number of samples to be in a leaf node is 1. Remaining parameters
are default for scikit-learn 0.19.1 version.

\subsection{Analysis based on strain rate minimum}

In order to estimate the failure time based on the time of minimum strain rate, $t_{\rm m}$, which separates the decelerating and accelerating creep regimes, we need to reconstruct a smooth signal from the discrete sequence of events. First, we note that the minimum of strain rate occurs during the linear creep regime and that during this regime plastic activity is almost exclusively thermally activated, with subsequent mechanically activated events being rare. In this case, avalanches have a typical size of a single plastic event and the strain increment measured over a certain observation interval is proportional to the number of avalanches occurring in that interval. Consequently, we can estimate the minimum of strain rate by looking for the maximum of a smoothed time increment signal. To obtain such smoothed signal, we substitute the $n$th value of the discrete time increments $\Delta t_n$ by the average of the increments whose numbers lie in the window $[n-h,n+h]$ of width $2h$ centered at $n$. Averaging over a window with a fixed width defined in terms of event number can be interpreted as averaging with an adaptative time window (i.e., a narrow time window in stages of small characteristic time increments and vice versa). The value $h$ of the window width must be set arbitrarily. We check the stability of the results upon variations of $h$ in order to decide its specific value. To this end, we compute the probability distribution of $t_{\rm m}$, where each $t_{\rm m}$ corresponds to a different realization of the creep process. We find that for a wide range of values $h \in [200,2000]$ the results are independent of $h$ for all the different values of the simulation parameters considered in this work, and we set $h=500$.



\begin{thebibliography}{99}
\bibitem{var_red}
The variance reduction criterion for splitting the data at any node in a random forest algorithm works as follows: 
If $N$ is the number of samples in a node, then the node is split (provided it is not restricted by maximum depth
of the tree) for a threshold value, say $A_0$ of a feature $A$, provided $A$ and $A_0$ maximize $\Delta$, which is given by
$\Delta=var(N)-\frac{n_l}{N}var(n_l)-\frac{n_r}{N}var(n_r)$, where $n_l$ and $n_r$ are the numbers of samples in the two nodes
if the split is accepted and $var(..)$ refers to the variance of the target variable i.e. the prediction time of the set of samples. 
This is checked for all features and all threshold values for all splitting decision. Still the individual trees are different simply because 
of the Bootstrapping described earlier.
\end{thebibliography}
\end{document}